\documentclass[a4paper,11pt]{article}
\pdfoutput=1 

\usepackage{jinstpub} 

\usepackage{lineno}
\linenumbers

\title{\boldmath KM3NeT Broadcast Optical Data Transport System}

\author[a]{S.~Aiello,}
\author[bd,b]{A.~Albert,}
\author[c]{S. Alves Garre,}
\author[d]{Z.~Aly,}
\author[e,f]{A. Ambrosone,}
\author[g]{F.~Ameli,}
\author[h]{M.~Andre,}
\author[i]{M.~Anghinolfi,}
\author[j]{M.~Anguita,}
\author[k]{M. Ardid,}
% ...

\affiliation[a]{INFN, Sezione di Catania, Via Santa Sofia 64, Catania, 95123 Italy}
\affiliation[b]{Universit{\'e}~de~Strasbourg,~CNRS,~IPHC~UMR~7178,~F-67000~Strasbourg,~France}
\affiliation[c]{IFIC - Instituto de F{\'\i}sica Corpuscular (CSIC - Universitat de Val{\`e}ncia), c/Catedr{\'a}tico Jos{\'e} Beltr{\'a}n, 2, 46980 Paterna, Valencia, Spain}
\affiliation[d]{Aix~Marseille~Univ,~CNRS/IN2P3,~CPPM,~Marseille,~France}
\affiliation[e]{INFN, Sezione di Napoli, Complesso Universitario di Monte S. Angelo, Via Cintia ed. G, Napoli, 80126 Italy}
\affiliation[f]{Universit{\`a} di Napoli ``Federico II'', Dip. Scienze Fisiche ``E. Pancini'', Complesso Universitario di Monte S. Angelo, Via Cintia ed. G, Napoli, 80126 Italy}
\affiliation[g]{INFN, Sezione di Roma, Piazzale Aldo Moro 2, Roma, 00185 Italy}
\affiliation[h]{Universitat Polit{\`e}cnica de Catalunya, Laboratori d'Aplicacions Bioac{\'u}stiques, Centre Tecnol{\`o}gic de Vilanova i la Geltr{\'u}, Avda. Rambla Exposici{\'o}, s/n, Vilanova i la Geltr{\'u}, 08800 Spain}
\affiliation[i]{INFN, Sezione di Genova, Via Dodecaneso 33, Genova, 16146 Italy}
\affiliation[j]{University of Granada, Dept.~of Computer Architecture and Technology/CITIC, 18071 Granada, Spain}
\affiliation[k]{Universitat Polit{\`e}cnica de Val{\`e}ncia, Instituto de Investigaci{\'o}n para la Gesti{\'o}n Integrada de las Zonas Costeras, C/ Paranimf, 1, Gandia, 46730 Spain}
% ...
\affiliation[bd]{Universit{\'e} de Haute Alsace, rue des Fr{\`e}res Lumi{\`e}re, 68093 Mulhouse Cedex, France}
% ...

\emailAdd{km3net-pc@km3net.de}

\abstract{The optical data transport system of the KM3NeT neutrino telescope at the bottom of the Mediterranean Sea will provide each of the more than 6000 optical modules in the detector arrays with a point-to-point optical connection to the control stations onshore. The ARCA and ORCA detectors of KM3NeT are being installed at a depth of about 3500 m and 2500 m, respectively; their distance to the control stations is about 100 kilometers and 40 kilometers. The expected maximum data rate is 200 Mbps per optical module. The implemented optical data transport system matches the layouts of the networks of electro-optical cables and junction boxes in the deep sea. For efficient use of the fibres in the system the technology of Dense Wavelength Division Multiplexing is applied. The performance of the optical system in terms of measured bit error rates, optical budget and the next steps in the implementation of the system are presented.}


\arxivnumber{} 

\collaboration[c]{on behalf of KM3NeT collaboration}

\begin{document}
\nolinenumbers
\maketitle
\flushbottom

\section{Introduction}
\label{sec:intro}
\begin{figure}[htbp]
\centering 
\includegraphics[width=.8\textwidth,origin=c]{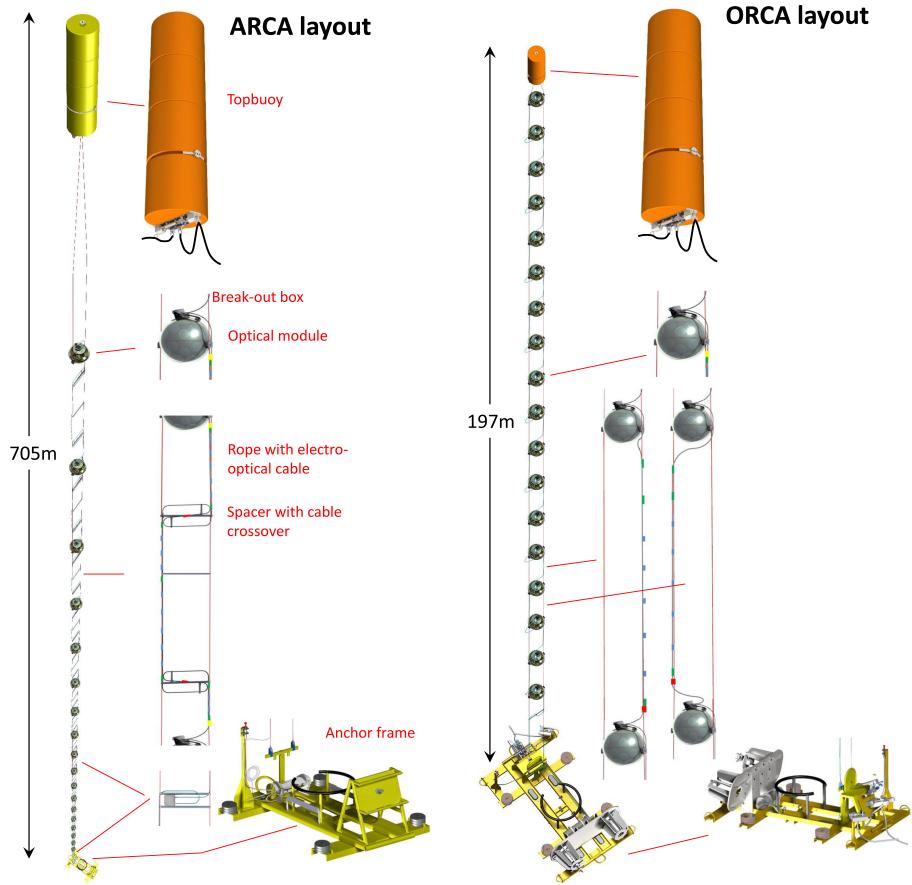}
\qquad
\caption{\label{fig:1} Layout of ARCA (left) and ORCA (right) detection unit showing the 18 optical modules on one detection unit, the electro-optical cable guided along the full length of the unit, as well as the top buoys and anchor frames \cite{c}.}
\end{figure}

KM3NeT is a network of neutrino telescopes in the Mediterranean Sea \cite{a}. The main scientific goals of KM3NeT are (i) neutrino astronomy, i.e. searching for cosmic neutrinos from distant astrophysical sources like gamma ray bursts, supernovae or coalescent stars and (ii) the study of oscillation patterns of atmospheric neutrinos to investigate the neutrino mass ordering. The first objective will be pursued with the KM3NeT/ARCA telescope in construction offshore Capo Passero, at about 3500 m under sea level, while the KM3NeT/ORCA detector, being installed at about 2500 m under sea level offshore the Toulon coast, in France, is mainly devoted to the second objective.
For the detection of neutrinos large three-dimensional arrays of about 2000 Digital Optical Modules (DOMs) each are being installed in the deep sea to detect the Cherenkov light emitted along the path of relativistic charged particles originating from the interaction of neutrinos with matter inside or in the vicinity of the arrays.
Each optical module is equipped with 31 photomultiplier tubes (PMTs) and calibration instrumentation together with all electronics for read-out and data acquisition \cite{b}. The modules in the arrays are integrated in Detection Units (DUs)- vertical structures anchored to the seabed; an electro-optical backbone cable running the full length (Fig.~\ref{fig:1}). At the anchor of a detection unit a base module provides the interconnection with the seafloor electro-optical cable network. The network connects the detection units with control stations onshore for data transfer and detector control.

\subsection{The detectors}
The configuration of the ARCA\footnote{\textit{Astroparticle Research with Cosmics in the Abyss}} detector is optimised for the detection of cosmic neutrinos with energies above about 1 TeV. When completed, it will consist of two blocks of 115 detection units each. The ARCA detection unit has a height of about 700 m and supports 18 optical modules (Fig.~\ref{fig:1} left). The vertical distance between optical modules is about 36 m, the horizontal distance between detection units is about 90 m. The instrumented volume of ARCA will be about one cubic kilometer. On the other hand, the configuration of the ORCA\footnote{\textit{Oscillation Research with Cosmics in the Abyss}} detector is optimised for the detection of atmospheric neutrinos with energies in the range 1 GeV-1 TeV. Compared to ARCA, the ORCA detector is relatively densely instrumented with optical modules and about a factor 100 smaller in volume. It comprises one block of 115 DUs about 200 m high; the vertical distance between the optical modules is about 9 m (Fig.~\ref{fig:1} right). With a horizontal distance between the detection units of about 20 m, the instrumented volume of ORCA correspond to about 7 megatonnes of sea water.

\subsection{The networks}
Data transmission and detector control of the arrays rely on networks of electro-optical cables and junction boxes which connect the optical modules in the deep sea to the control stations onshore. 

\begin{figure}[htbp]
\centering 
\includegraphics[width=.8\textwidth,origin=c]{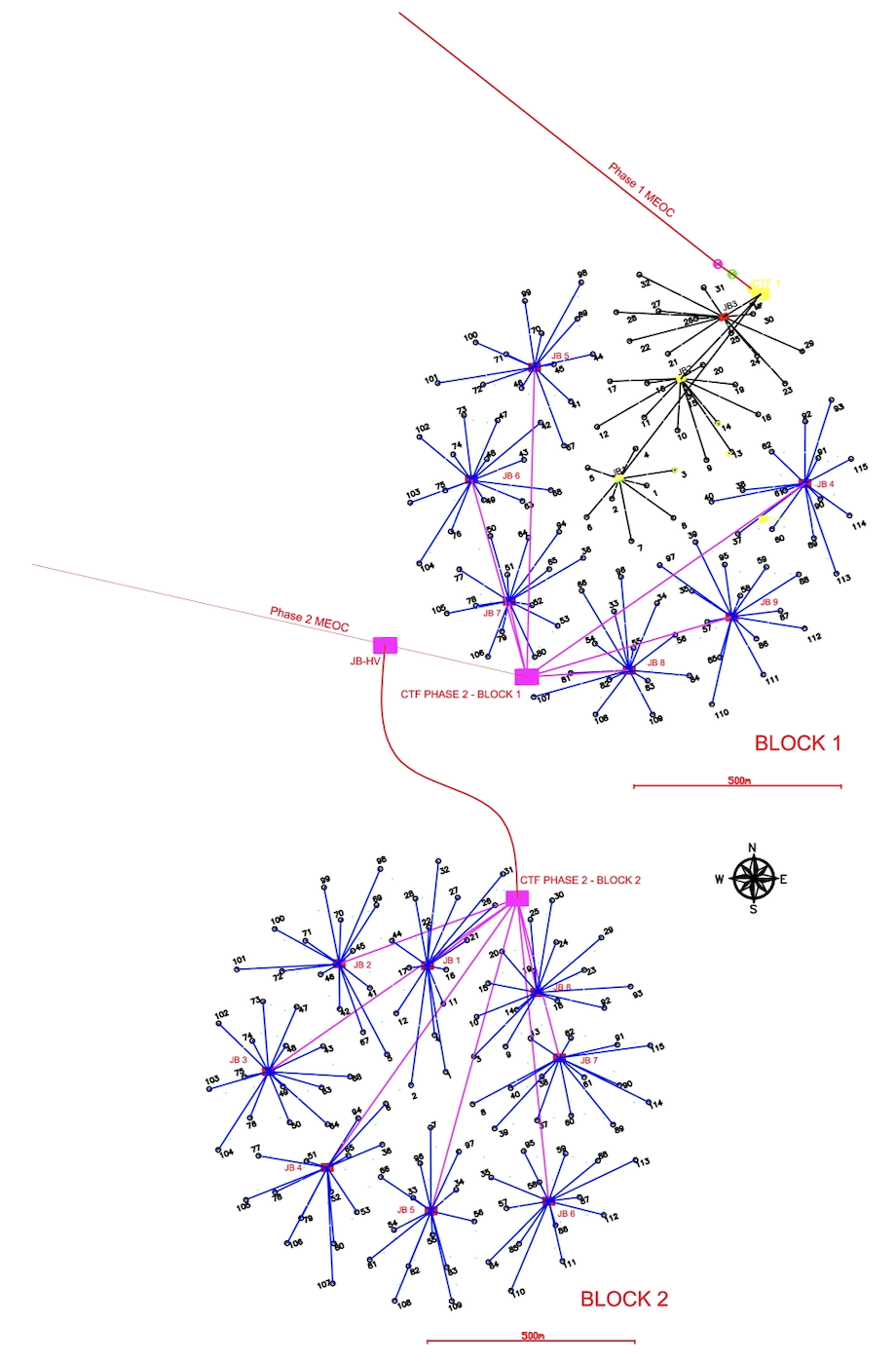}
\qquad
\caption{\label{fig:2} Layout of the ARCA network. Boxes represents the junction boxes; Dots indicates the detection units connected to the junction boxes. The junction boxes are connected to the cable termination frames of the Main Electro-Optical Cables (MEOCs) to shore (red lines to the pink boxes). The black lines reflect the implementation of the network in the first phase of KM3NeT.}
\end{figure}

The detection units of ARCA will be connected to the deep-sea network infrastructure being deployed about 100 kilometres off the coast of Sicily near Portopalo di Capo Passero in Italy (Fig.~\ref{fig:2}). The network has a star configuration that provides the required horizontal distance of about 90 m between the anchored detection units. The configuration minimises the complexity of the cabling and eventually the cost of the detector installation.
Up to twelve detection units are connected to a hub, the junction box; junction boxes are then connected to the termination frame of the main electro-optical cables and finally to the control station. Each detection unit has a vertical backbone cable with 18 fibres for individual connection between each optical module and the base of the DU; the cable connecting the base module of the detection unit to a junction box has two fibres; the cable between a junction box and the main cable termination frame has four fibres; for historical reasons the first main cable to shore has 20 optical fibres; the second one has 48 fibres.
The ARCA broadcast architecture foresees a total of three junction boxes for the first phase of KM3NeT.

\begin{figure}[htbp]
\centering 
\includegraphics[width=.8\textwidth,origin=c]{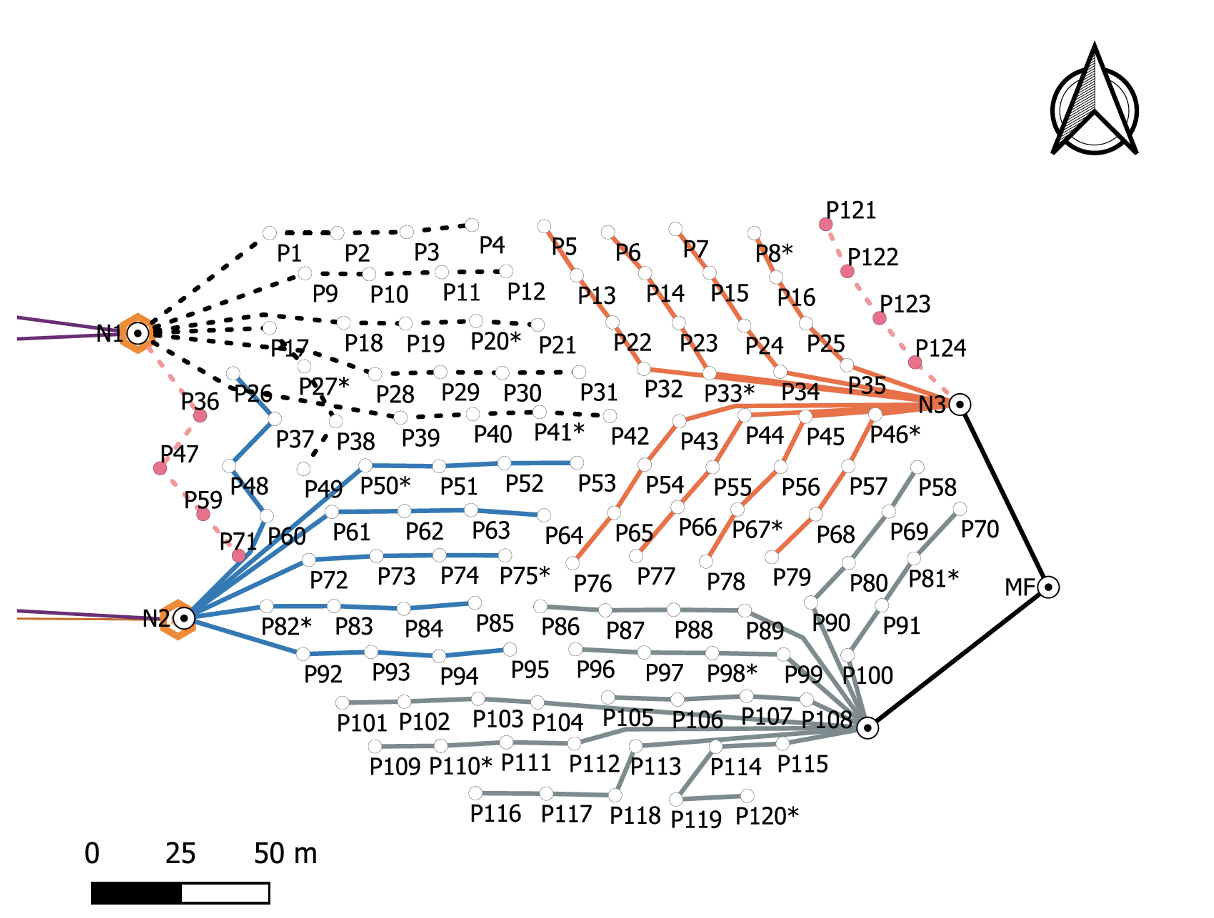}
\qquad
\caption{\label{fig:3} Layout of the ORCA network. The node N1 and N2 are daisy-chained in the main electro-optical cable and reflect the implementation of the network in the first phase of KM3NeT. Dots are detection units daisy-chained to a node.}
\end{figure}

In parallel, a network infrastructure is being deployed about 40 kilometres off the coast of Toulon in France (Fig.~\ref{fig:3}) for the ORCA detector. Its design foresees groups of four detection units connected to each other in daisy chain. Each daisy chain is connected to a hub - the node. The horizontal distance between the detection units is about 20 m; up to 24 detection units can be connected to each node. 
The nodes are then connected in daisy chain with the manifold at the end of the main electro-optical cables to the control station onshore. 
A diverse number of fibres are used in different sections of the network: the vertical backbone cable in a detection unit has 18 fibres; the cable in the daisy chains connecting four base modules has 4 fibres; the cables of the daisy chain of nodes comprise 36 optical fibres for the first main cable to the control station; the second cable is the one used in the ANTARES telescope \cite{d} has 48 fibres and will be used for the second phase of construction of the ORCA telescope.
(De)multiplexing the optical data transmission channels and operating the various detector control channels of the fibre networks are the main tasks of the optical data transmission system described in the next sections. Details of the physical layer of the optical data transfer and detector control system are presented.

\section{The KM3NeT optical data transfer system}
Because of the large distances between the DUs and the control station and the expected average data rate of 30 Mbps per optical module with a maximum of 200 Mbps \cite{e}, the data transmission system of KM3NeT has to rely on an optical fibre physical layer. To minimise the number of fibres in the underwater electro-optical cables, Dense Wavelength Division Multiplexing (DWDM) is applied in order to share the same fibre between several optical modules \cite{f}.
The vertical electro-optical backbone cable in the DU routes a single fibre from each optical module to the base module, which is located at the anchor of the DU. Each optical module is connected to the system of the control station onshore through its own point-to-point optical link. In order to achieve this, a different frequency is assigned to each optical module and allocated on a fixed grid with spacing of 200 GHz between optical modules of the same detection unit. According to the frequency allocation, four different channel group configurations for the detection units are defined, labeled A, B, C and D. The frequency grids of DOMs hosted on neighbouring DUs in a chain of four have an offset of 50 GHz.

\subsection{Multiplexing}
A multi-stage multiplexing scheme is adopted offshore. The first stage is performed in the base module of each detection unit, the second stage in the junction box in the case of ARCA or in the first detection unit of a daisy chain for ORCA. With this scheme it is eventually possible to multiplex the optical modules of four detection units with a total of 72 different frequency channels using a single fibre, with a spectral spacing of 50 GHz \cite{g}.
Optical modules and base modules house a Central Logic Board (CLB) \cite{h} which collects all information produced by the PMTs and the other instruments and transmits it to the control station through the assigned DWDM channel. The CLB hosted in the DOM collects also data from the ancillary instrumentation of the DOM: piezo sensor, compass and nanobeacon.
Clock information, slow control and detector control commands are broadcasted by the data acquisition (DAQ) system in the control station and used to manage subsystem functions inside the optical modules and the base modules. The electrical to optical conversion of the various signals is implemented by means of Small Form-factor Pluggable (SFP) optical transceiver modules \cite{i} hosted in the CLB. The CLB in the base module also manages the ancillary instrumentation of the detection unit: hydrophone and long base line acoustic beacon.

\subsection{Bandwidth}
The bandwidth usage is asymmetric: large in the upstream direction, from the DOMs to the onshore DAQ system, and small in the downstream direction from the onshore DAQ system to the DOMs and base modules. The main contributor to the optical module upstream data rate is the PMT signal acquisition \cite{l}, which accounts for sustained rates between 20 and 100 Mbps, with seconds-long bursts with rates up to five times higher. A further contribution comes from the piezo detectors inside the optical modules used for acoustic positioning, accounting for 10 Mbps. All remaining instrumentation together produces less than a few hundred of Kbps. The bandwidth asymmetry is reflected in the design of the system. All serial streams have the same data rate of 1.25 Gbps Ethernet. The downstream slow control and the detector control signals are broadcasted towards all optical modules and base modules connected to each junction box or to a node – a point to multi-point configuration. The Ethernet stream bandwidth is shared among various detection units. Each optical module and base module exploits the full Ethernet bandwidth point to point for the upstream communication.

\subsection{Synchronisation}
Since neutrino event reconstruction is based on the PMT signal time, a common timing reference must be available to the CLBs to allow for detector wide synchronisation. The time offset between each acquisition channel and the fixed reference must be known to compare signal times. In order to facilitate the clock distribution, the White Rabbit (WR) protocol over Ethernet is used \cite{m}. Synchronisation is achieved by communication between a WR master onshore, realized by means of a specific WR switch (WRS) fabric, and a WR slave unit hosted on the CLB in the base module offshore. This makes it possible to dynamically reconstruct a time offset between downstream and upstream clocks.
In this way, all the receivers are synchronised by design to the onshore time reference, which is derived from a GPS station \cite{n}.
In particular, the above mentioned asymmetric broadcast design implies a modification in the onshore WRS fabric with respect to the standard implementation.
The downstream communication is handled by the WRS called Broadcast. The upstream reception is implemented in the WRS called Level 1 for the base channels of the DU and in a standard switch fabric for the DOMs. The WRS called Bridge, connected to the GPS, acts as the Grand Master and distributes the timing to the rest of the WRS fabric.

\section{Implementation for ARCA}
\begin{figure}[htbp]
\centering 
\includegraphics[width=.9\textwidth,origin=c]{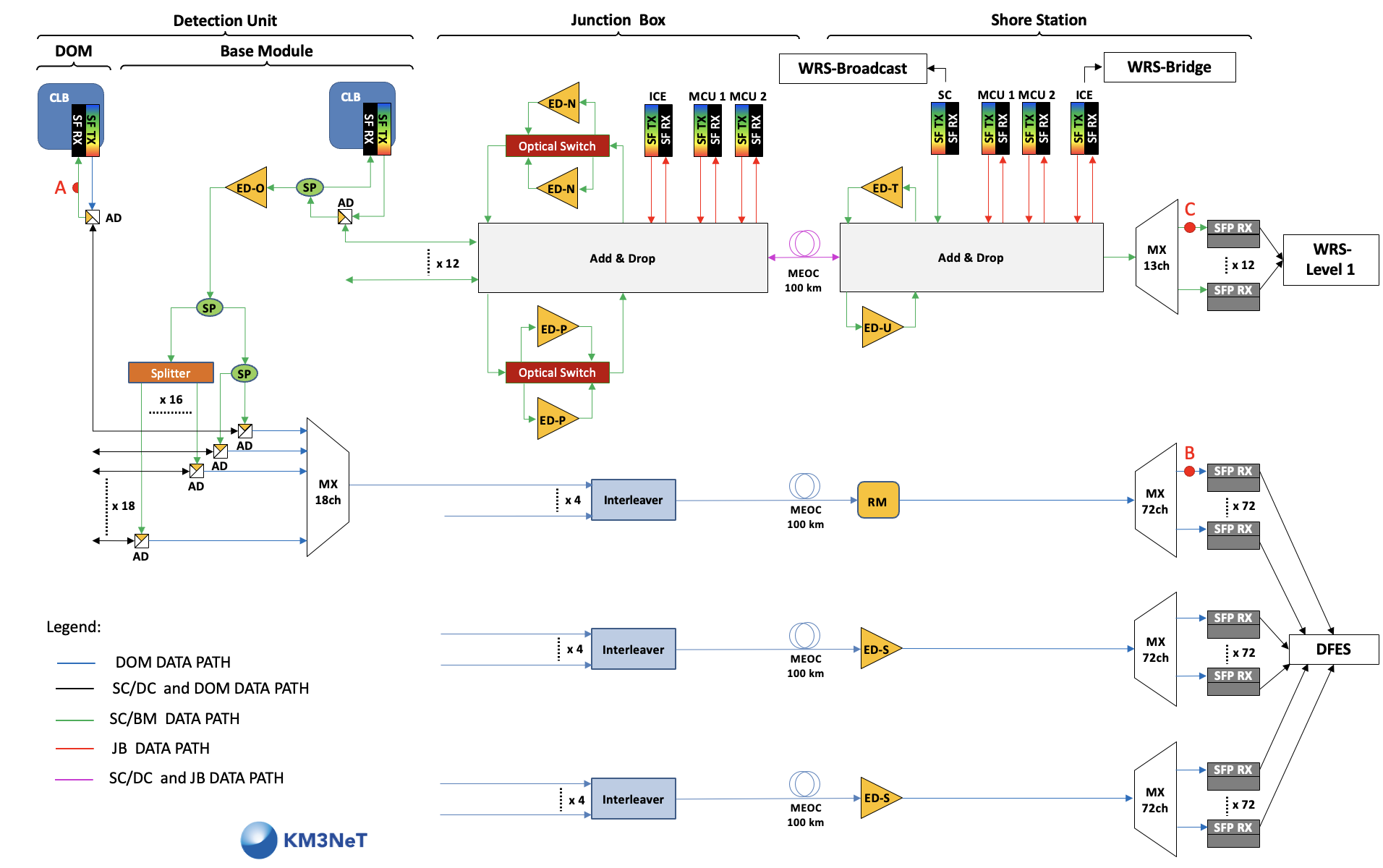}
\qquad
\caption{\label{fig:4} The ARCA optical system. The colored lines represent the different optical paths such as Digital Optical Module (DOM), Slow Control (SC), Base Module (BM) and Junction Box (JB) data. Details are described in the text. The red dots A, B and C indicate different measuring points that are explained in section 5.}
\end{figure}

The ARCA optical system, shown in Fig.~\ref{fig:4}, matches the seafloor electro-optical infrastructure with a star topology (Fig.~\ref{fig:2}). In the first phase of construction up to twelve DUs are connected to one junction box by means of two fibres per detection unit and each junction box is connected to shore using four fibres of the main electro-optical cable. The ARCA optical system is composed of three main optical paths:

- a channel for slow control (downstream) and a base module channel (upstream) for each detection unit;

- three channels for the junction box control and command (downstream-upstream);

- a channel for each optical module data (upstream).

The slow and the base module channels share the same fibre bidirectionally with the channels for junction box control and command. The DOM data channel uses a dedicated fibre; only in the vertical electro-optical backbone cable in the detection unit fibres are shared with the slow control channel and with the base modules. The slow control channel is broadcasted by the SFP transceiver hosted in the WRS-Broadcast, and then multiplexed by add-drop filters with the primary and secondary (backup) channels for junction box control and command and the junction box instrumentation channel hosted in the WRS-Bridge. The four channels are routed to a booster Erbium Doped Fiber Amplifier (EDFA, ED-T in Fig.~\ref{fig:4}) the output of which is routed to a circulator. The circulator bidirectional port is then connected to the main electro-optical cable.

At the junction box, the channels for control and command, for instrumentation control and the channel for slow control are demultiplexed. The junction box control and command channels are routed to the monitor and control unit, the instrumentation control channel is routed to the instrumentation control electronics boards, and the slow control channel is routed to an EDFA module (ED-N in Fig.~\ref{fig:4}).
The ED-N boosts the slow control channel before it is added again to the bidirectional fibre, split with a 1:16 ratio and eventually routed to each of the 12 detection units connected to the junction box\footnote{Splitters with 1:16 ratio have been used as 1:12 ratio splitters are not available on the market}.
At the base module, the slow control channel is dropped by an add-drop filter and then split: 60\% is routed to the CLB of the base module and 40\% amplified again by an EDFA hosted in the base module (ED-O in Fig.~\ref{fig:4}).
The base module data channels are transmitted through SFP transceivers hosted in the CLBs of the base modules and added to the slow and detector control segment by an add-drop filter. At the junction box, the same 1:16 ratio splitter used for the slow control channel is used to combine up to 12 base module channels. The combo of base module channels must be amplified at the junction box before being routed to the main electro-optical cable. For this reason, the base module data channels are dropped and routed to an EDFA module (ED-P in Fig.~\ref{fig:4}). After amplification, the base module channels are added again to the bidirectional fibre and multiplexed with the channel for junction box control and command and with the channel for instrumentation control.
For each primary EDFA in the junction box a backup EDFA is added to comply with the reliability studies carried out to maintain a life span of 20 years of each junction box. An optical switchover logic board is introduced, which switches from the main amplifier to its backup in case of a failure of a main EDFA. An optical switchover board is dedicated to each amplifier pair (ED-N and ED-P in Fig.~\ref{fig:4}). The board is controlled remotely via the junction box control and command channels. At the control station, a circulator separates the upstream and downstream directions, followed by a cascade of add-drop filters which drop the channel for junction box control and command and the channel for instrumentation control from the base module data channels combo. The former are routed to the junction box control and command unit, the latter to the WRS-Bridge. The remaining base module channels are amplified by an EDFA (ED-U in Fig.~\ref{fig:4}), routed to a demultiplexer (MX 13ch) and eventually passed to Avalanche PhotoDiode (APD) SFP receivers hosted in WRS-Level 1.
The data acquired by the optical modules are transmitted via SFP transceivers hosted in the CLBs of the modules and added by filters to the channel for slow control and base module data. The optical module channels then share the vertical electro-optical cable fibre with the slow control channel. At the base module of the detection unit they are dropped by 18 add-drop filters and routed to an 18 channel multiplexer (MX 18ch). At the junction box four sets of 18 channels, corresponding to the four detection unit configurations -- labelled type A, B, C and D -- are routed to a 200-50 GHz interleaver. At the control station, the optical module channels are routed to an EDFA S or a Raman two pumps amplifier (ED-S or RM-2 in Fig.~\ref{fig:4}), to a 72 channel demultiplexer (MX 72ch) and eventually routed to APD SFP receivers hosted in the optical module front-end switches (DFES in Fig.~\ref{fig:4}) of the optical modules. The choice of EDFA amplifier or Raman amplifier is driven by market availability.

\section{Implementation for ORCA}
\begin{figure}[htbp]
\centering 
\includegraphics[width=.9\textwidth,origin=c]{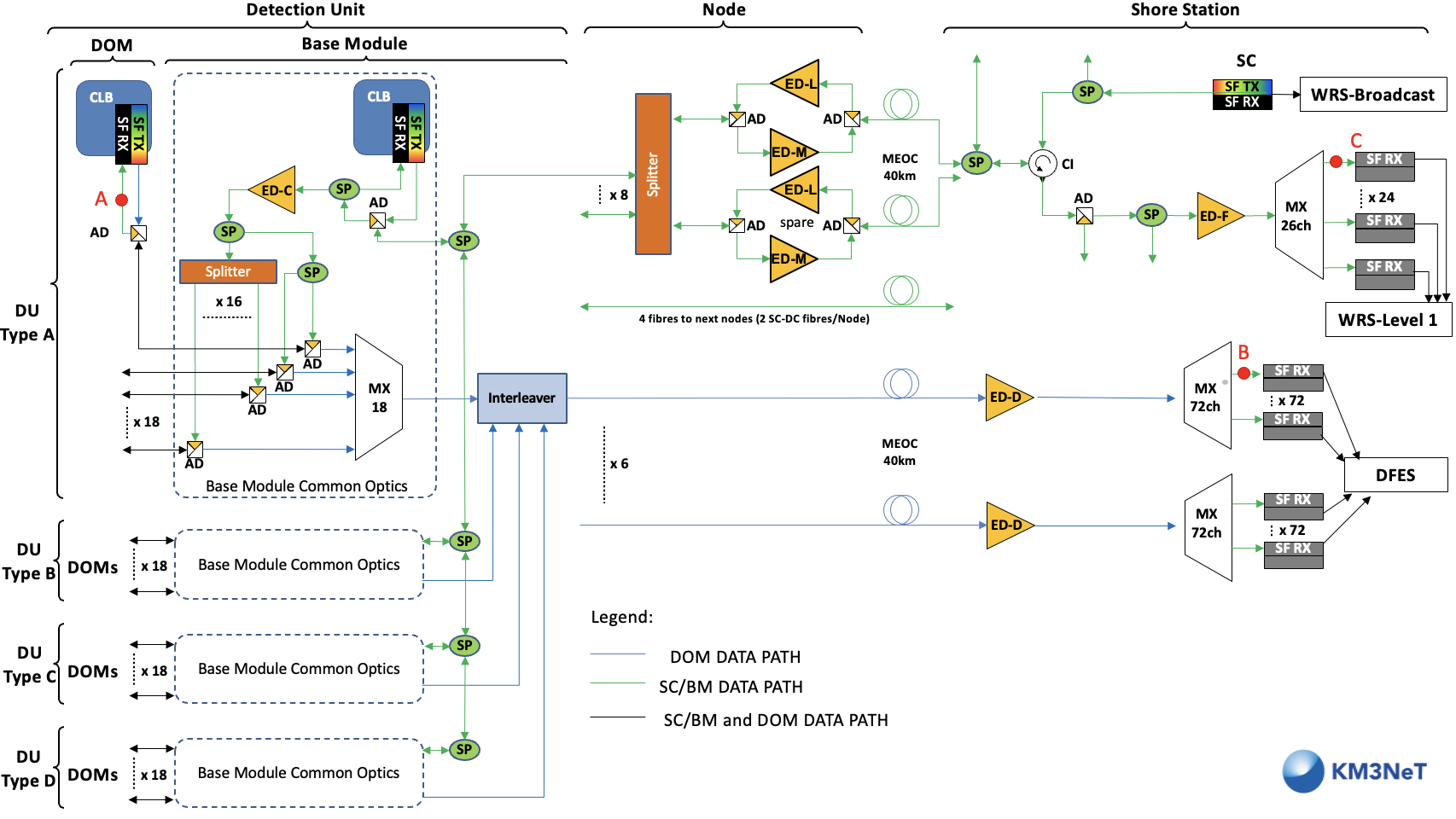}
\qquad
\caption{\label{fig:5} The ORCA optical system. The colored lines represent the different optical paths such as Digital Optical Module (DOM), Slow Control (SC) and Base Module (BM) data. Details are described in the text. The red dots A, B and C indicate different measuring points that are explained in section 5.}
\end{figure}

The ORCA system, shown in Fig.~\ref{fig:5}, is matched to the seafloor electro-optical cabling infrastructure with a daisy chain topology: groups of four different detection units (type A, B, C and D, each one using a set of 18 channels) are connected to a central hub called node. The node can provide connectivity for up to 24 detection units and for this purpose requires six fibres of the main electro-optical cable (two additional fibres are used as spare). Two nodes are then connected in cascade exploiting the full main cable fibre capacity. The system is composed of three channels:

- one channel for the slow control (downstream);

- one channel for the base module data (downstream-upstream);

- one channel for the optical module data (upstream).

The ORCA daisy chain topology foresees in different splitter ratios between base modules along the slow control and base module data optical layers in order to equalise the optical power levels received by the base modules, as shown in Fig.~\ref{fig:5}.
The first detection unit of the chain (type A) houses the interleaver, which multiplexes the optical module data produced by the four detection units in the daisy chain. The slow control and base module channels share the same fibre bidirectionally. The optical module data channel uses a separate set of fibres, with the exception of the vertical electro-optical cable fibres that are shared only with the slow control channel. The slow control and base module channel interconnects the onshore DAQ system with the optical modules and the base modules data.
The slow control channel is broadcasted by the SFP transceiver hosted in the WRS-Broadcast, and routed to the input port of the circulator. The circulator bidirectional port is then connected to the main electro-optical cable. At the node, the slow control channel is demultiplexed from the bidirectional fibre and routed to an EDFA module (ED-L in Fig.~\ref{fig:5}).
The ED-L boosts the slow control channel before it is added again to the bidirectional fibre, split with a 1:8 ratio and eventually routed to each of the eight node outputs connected to six detection units daisy chains (two out of eight are kept as spare). Inside the base modules of the detection units the slow control channel is dropped by an add-drop filter and then split: 60\% is routed to the CLB in the base module and 40\% is amplified by an EDFA module (ED-C in Fig.~\ref{fig:5}). The slow control channel is routed to each of the 18 optical modules, by a combination of 1:16 and 1:2 ratio splitters. The access to the vertical electro-optical cable fibres is implemented by add-drop filters housed in the base module and in each DOM. The base module data channels are transmitted from the SFP transceivers hosted in the CLB of the base modules and added to the slow control and base module channel through an add-drop filter. At the node, the 1:8 ratio splitter is used to combine 24 base module data channels which are amplified before being routed to the main cable. For this reason, the base module data channels are routed to an EDFA module (ED-M in Fig.~\ref{fig:5}). After amplification, the base module channels are added to the bidirectional fibre. Similarly to ARCA, for each main EDFA in the node, a backup EDFA is added. In the case of ORCA, the redundancy scheme is implemented exploiting the second input port of the 1:8 ratio splitter. The backup EDFAs are enabled by manually connecting the shore end point to the spare node fibre. At the control station onshore, the base module data channel is routed to the 26 channel demultiplexer (MX 26ch) and eventually to the APD SFP receiver hosted in the WRS-Level 1. In contrast to ARCA, the ORCA system node control and command does not share the same fibre of the slow control and base module channel; instead it is routed via two separate fibres which are shared with the other nodes.
An additional set of two fibres in the main cable to shore is dedicated to instrumentation for research in Earth and Sea sciences, which is part of the KM3NeT infrastructure. The data from the optical module channels are transmitted by SFP transceivers hosted in the CLBs of the optical modules and added to the slow control and base module data segment by one add-drop filter. At the base module they are demultiplexed from the slow control and base module channel by 18 add-drop filters and routed to an 18 channel multiplexer (MX 18ch). At the control station the DOMs channels are routed to an EDFA (ED-D in Fig.~\ref{fig:5}), to an 72 channel demultiplexer (MX 72ch) and eventually to the APD SFP receivers hosted in the front end switches (DFES in Fig.~\ref{fig:5}).

\section{Performance of the system}
In order to assess the optical performance of the ARCA and ORCA implementations, a key performance indicator is the ratio of the number of bit errors over the total number of transferred bits during a certain time window. This ratio is also known as Bit Error Rate (BER). A laboratory testbed was setup to measure the BER at several locations in the optical networks.

\subsection{The BER test setup}
\begin{figure}[htbp]
\centering 
\includegraphics[width=.8\textwidth,origin=c]{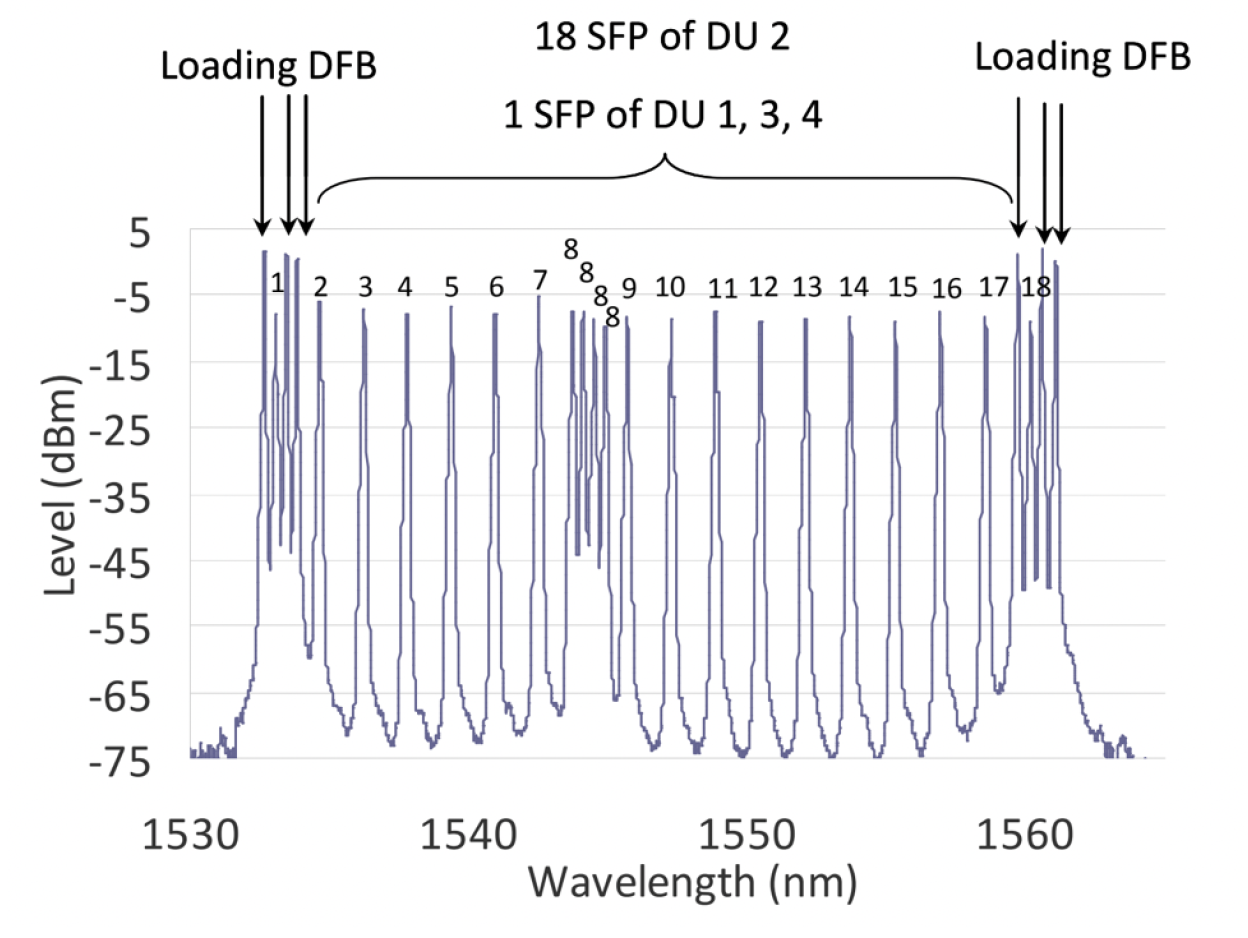}
\qquad
\caption{\label{fig:6} Optical spectrum of detection unit channels (corresponding to each optical module) in the test setup.}
\end{figure}

The designs of the optical networks of ARCA and ORCA in Fig.~\ref{fig:4} and Fig.~\ref{fig:5} respectively was implemented in a laboratory test setup. Some parts of the network, such as optical fibres of the vertical electro-optical cable, the main electro-optical cable and the seafloor cabling were emulated by optical attenuators, allowing for a compact testbed. In the detector, the SFP transceivers are hosted in the CLBs and DAQ switches onshore, while in the testbed they were driven at 1.25 Gbps by one single Ethernet switch. The Ethernet switch hosted additional SFP transceivers to load the system; these were not monitored in the BER measurements. The SFP transceivers used for the BER measurement are hosted in several evaluation boards that drove them with a pseudo random binary sequence of $(2^{15}-1)$ bit using a non-return-to-zero line code.
In order to measure the BER of the optical channel under test, the received electrical signal produced by the evaluation board was compared with the transmitted. The bit-to-bit comparison of the two serial streams was implemented by the BER tester equipment. For each optical channel under test, several measurements were performed. Starting from the typically expected received power, the signal level was gradually lowered using a variable optical attenuator, placed in front of the receiver, until the level was so low that the link was lost (asserting the loss of signal flag).
From that point onwards, the optical signal level was increased again in small steps of 0.5 or 1 dB until the link was established and the errors on the received bit stream were counted. Aim of the test was to check the quality of the signals belonging to the optical module, slow control and base module data channels, with particular focus on verifying the amount of possible crosstalk between them, especially in the bidirectional segments of the optical system. To test the path of data from the DOMs, a dedicated configuration was designed to optimise electronic hardware that could reproduce the optical network. Only some channels belonging to a certain detection unit were investigated, and due to the port limitation of the Ethernet switch not all the neighbouring channels were installed. The BER was measured for the following channels in a detection unit (Fig.~\ref{fig:6}):

- channel 1: to check if there is a penalty (a degradation of the receiver sensitivity) due to the edges of the EDFA passband;

- channel 4: the least powerful channel;

- channel 7: the most powerful channel;

- channel 8: a channel with several neighbouring channels 50 GHz apart;

- channel 18: to check if there is a penalty due to the edges of the EDFA passband.

An example of the optical spectrum of the data channel from the optical modules is shown in Fig.~\ref{fig:6} where in addition to the DU channels, six continuous wave (CW) laser sources are introduced. The CW laser sources, labelled loading Distributed FeedBack (DFB) are located at the edges of the spectrum as neighbours of channel 1 and channel 18 with a separation of 50 GHz (0.4 nm) from the channels. The power level of the CW laser sources is similar to the power level of the signals belonging to the other detection units not present in the spectrum, so that the total optical power is equal to what is expected by the signals belonging to a set of A, B, C, D type of detection units. Eighteen channels belonging to the detection unit B type frequency grid are shown in Fig.~\ref{fig:6}, with channel 8 surrounded by channel 8 of type A, C and D detection units.
Using the testbed, the BER was measured for each optical segment in the ARCA and ORCA optical systems. The SFP transceiver module used for the BER measurements was the OE Solution model RDP12SZXSxxC/H, which has a declared sensitivity of $-25$ dBm at the beginning of life ($-26$ dBm at the end of life) and a minimum loss of signal assert level of $-35$ dBm. The measurements show that the actual sensitivity of the SFP transceiver is better than $-31$ dBm for a BER of $10^{-12}$ in the worst case, due to the effect of the EDFA preamplifiers before the receivers input. Along with the BER, the reliability of the measurement was tested by computing the $p$-value assuming a binomial distribution of errors \cite{o}:
$$p{\rm-value} =  1-\sum_{k=0}^{N-1}\Bigg(\frac{n!} {k!(n-k)!}\Bigg)p^k_h(1-p_h)^{n-k}$$
where $p_h$ is the value of BER, $n$ is the total number of bits transmitted during the measurement window and $N$ is the total number of bit errors. The smallest $p$-value is larger than 90\%, which well certifies the compatibility of the measurement with the assumption.

\section{Results}
\subsection{The slow control channel}
\begin{figure}[!htb]
   \begin{minipage}{0.48\textwidth}
     \centering
     \includegraphics[width=1\linewidth]{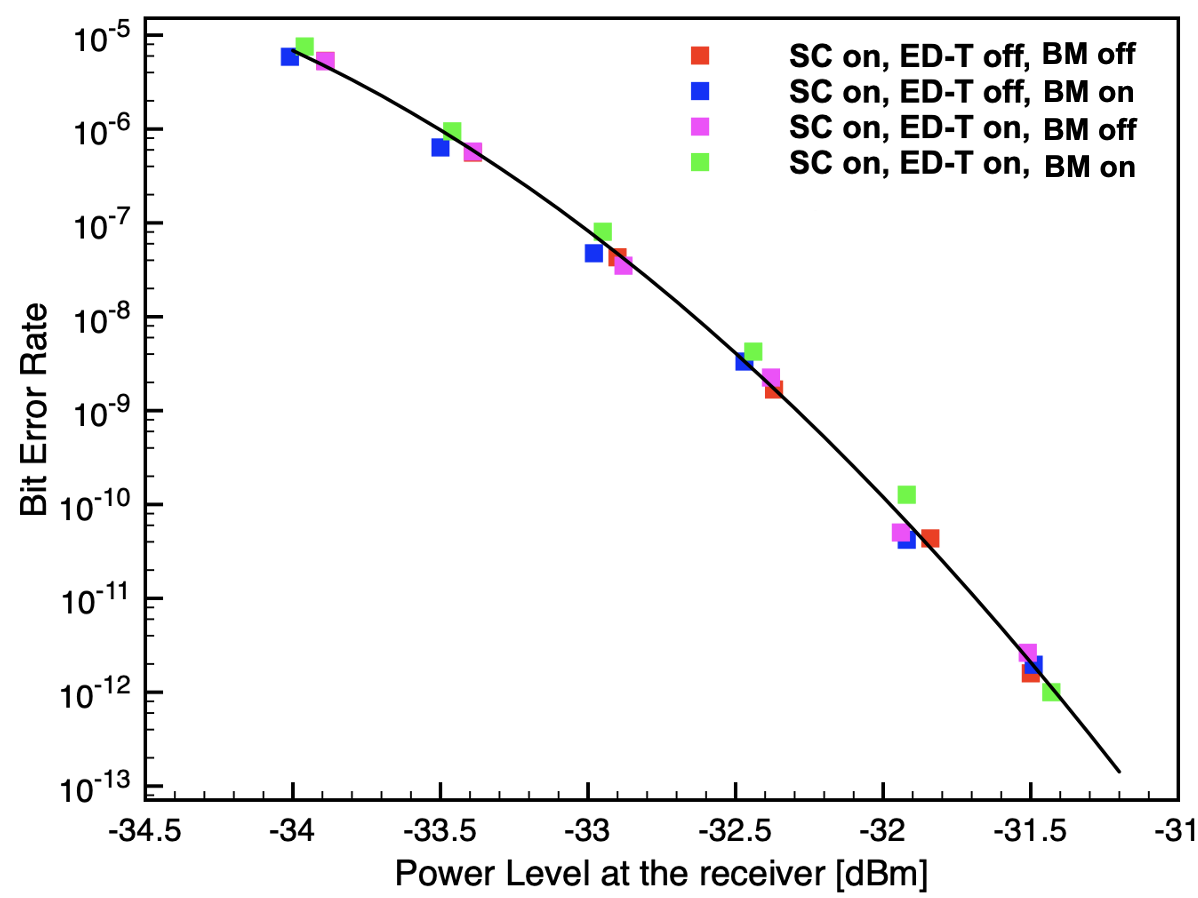}
     \caption{\label{fig:7} ARCA Bit Error Rate as function of power level of the receiver for the slow control channel offshore. Black line is drawn just to guide the eye.}
   \end{minipage}\hfill
   \begin{minipage}{0.48\textwidth}
     \centering
     \includegraphics[width=1\linewidth]{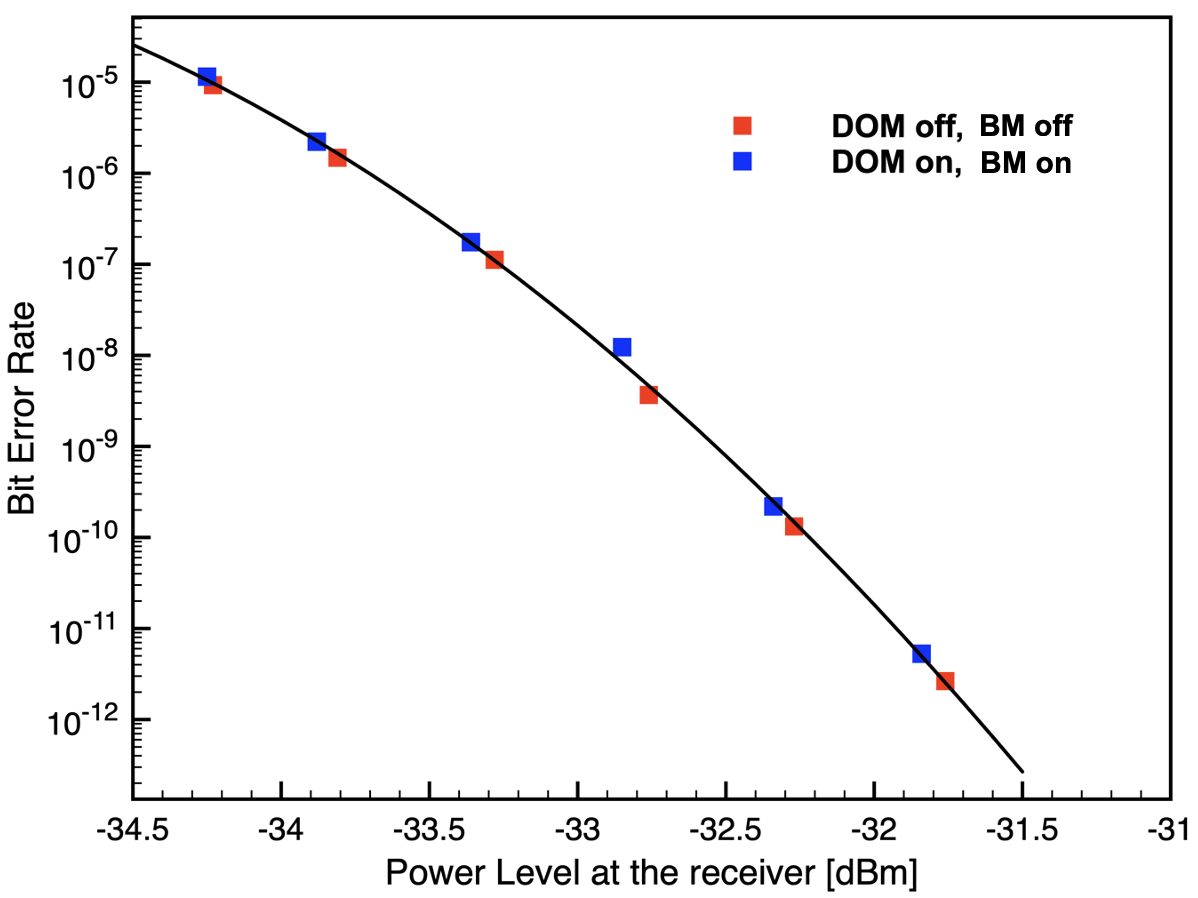}
     \caption{\label{fig:8} ORCA Bit Error Rate as function of power level of the receiver for the slow control channel offshore. Black line is drawn just to guide the eye.}
   \end{minipage}
\end{figure}

The BER of the slow control channel was measured by the receiver at a location near the CLB at point \textbf{A} in Fig.~\ref{fig:4} and Fig.~\ref{fig:5}.
The measurements were implemented with various combinations of slow control signals with or without the detector control channel and with or without ED-T amplification. In the ARCA testbed, ED-T operates at a 10 dB automatic gain mode. The resulting BER curve in Fig.~\ref{fig:7} show no differences for all considered cases indicating that the EDFA gain does not lead to a degradation of the receiver sensitivity. In addition, the presence of the detector control signal did not have any influence on the performance.

In the case of the ORCA architecture, the BER of the slow control link was measured with and without DOM and base module data, as shown in Fig.~\ref{fig:8}. Aim of this measurement was to determine if any crosstalk between the optical modules and slow control channel or base module and slow control channel was present, since they share the same fibres in the vertical electro-optical cable and in the fibre between the base module and the shore DAQ system respectively, as shown in Fig.~\ref{fig:4} and Fig.~\ref{fig:5}.

\subsection{The base module data channel}
\begin{figure}[!htb]
   \begin{minipage}{0.48\textwidth}
     \centering
     \includegraphics[width=1\linewidth]{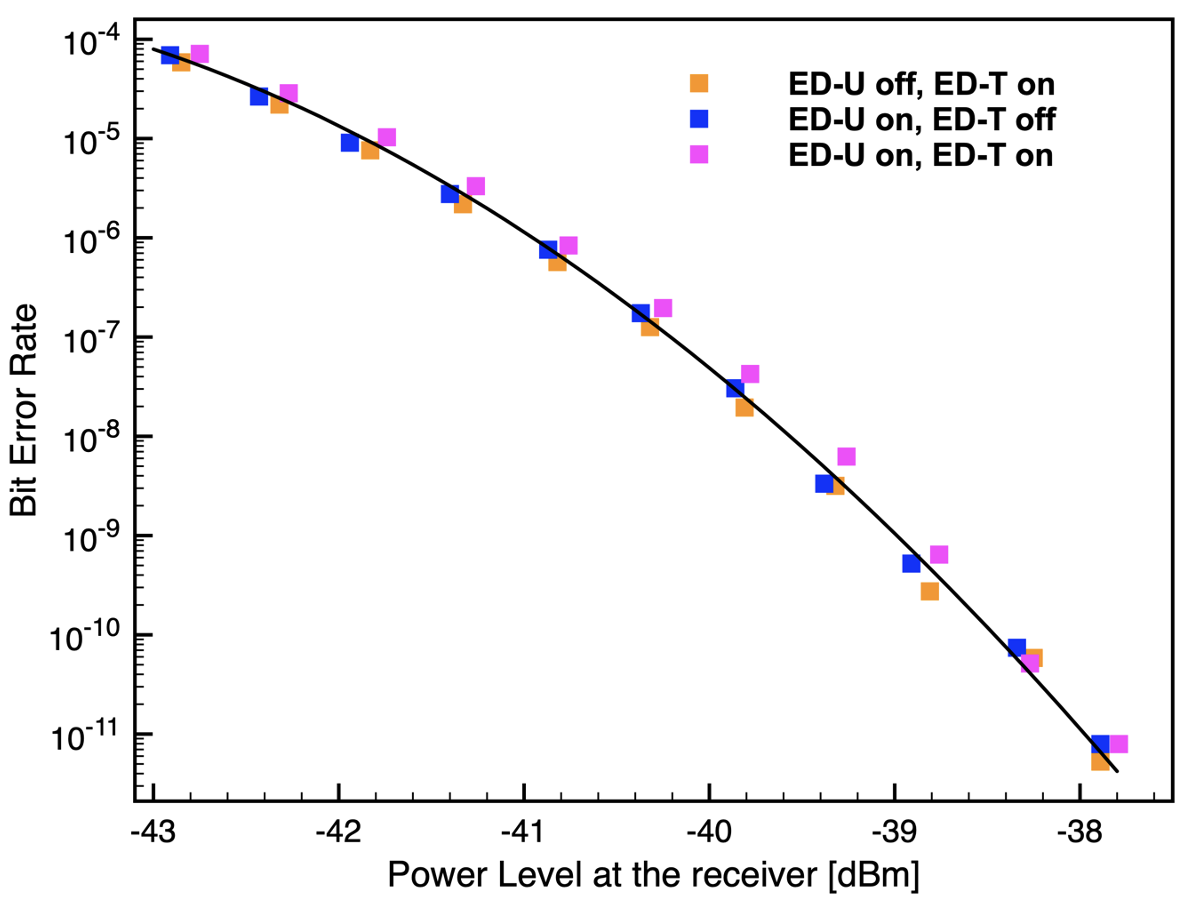}
     \caption{\label{fig:9} ARCA Bit Error Rate as function of power level of the receiver for the base module channel onshore. Black line is drawn just to guide the eye.}
   \end{minipage}\hfill
   \begin{minipage}{0.48\textwidth}
     \centering
     \includegraphics[width=1\linewidth]{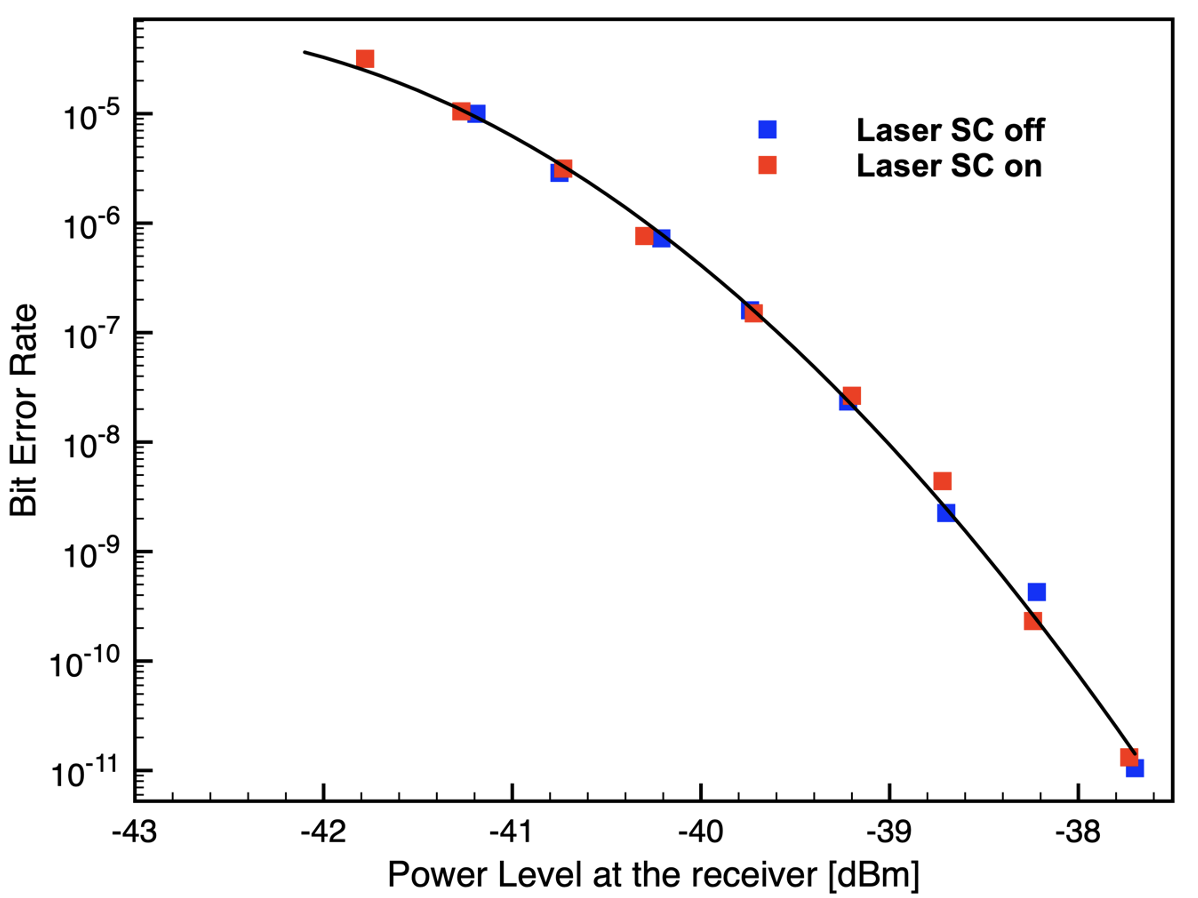}
     \caption{\label{fig:10} ORCA Bit Error Rate as function of power level of the receiver for the base module channel onshore. Black line is drawn just to guide the eye.}
   \end{minipage}
\end{figure}

For the base module data channel the BER was measured with the receiver in the shore station at point \textbf{C} in Fig.~\ref{fig:4} and Fig.~\ref{fig:5}. The results are shown in Fig.~\ref{fig:9} and Fig.~\ref{fig:10}.

The influence of the DOM data channels was not considered in this case as these channels do not share the base module fibre. For ARCA the bit error rate curve at the shore station receiver with and without ED-T emission, is shown in Fig.~\ref{fig:9}. The data indicate that the addition of ED-U amplification brings no signal degradation. The BER curve for the ORCA base module channel is shown in Fig.~\ref{fig:10}. The curves show no influence of the slow control channel on the base module channel even though they share the same optical path.

\subsection{The optical module data channel}
\begin{figure}[!htb]
   \begin{minipage}{0.48\textwidth}
     \centering
     \includegraphics[width=1\linewidth]{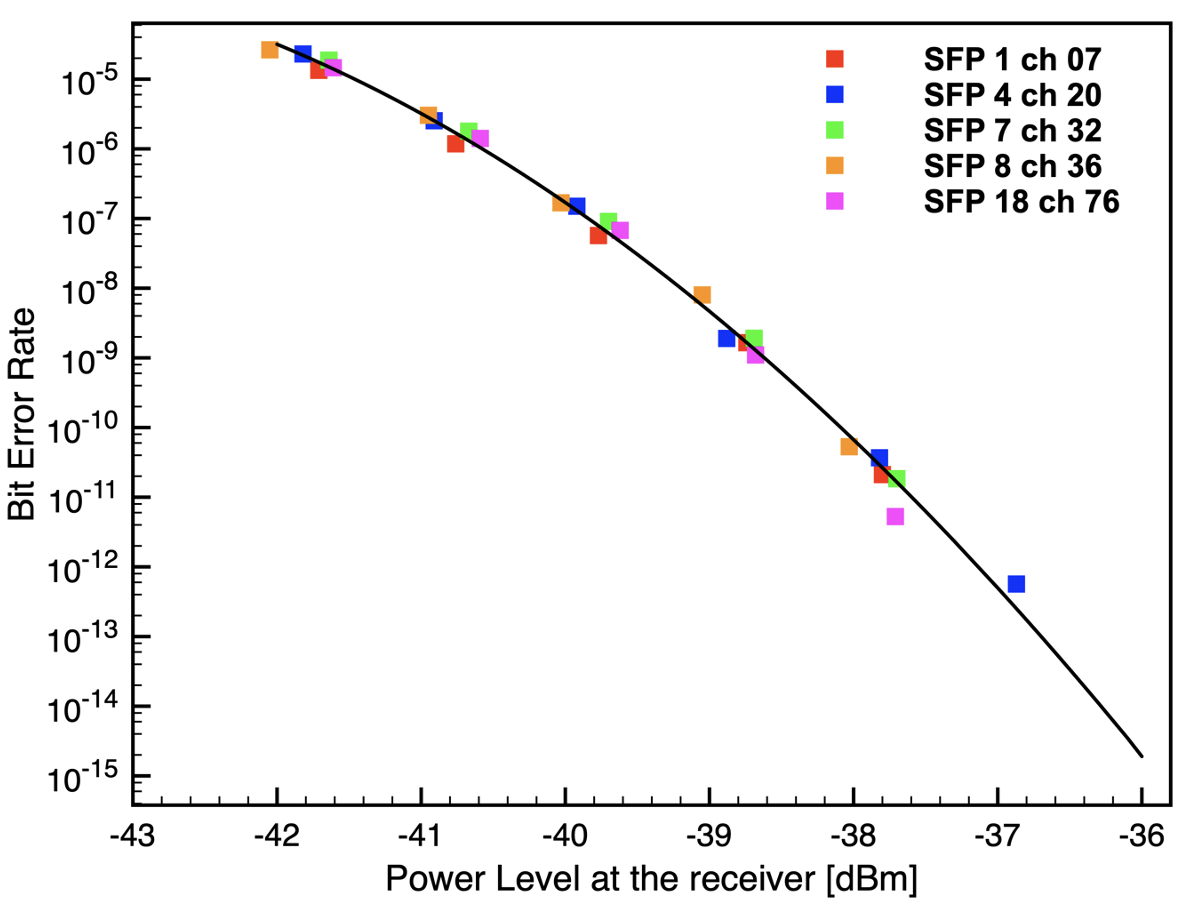}
     \caption{\label{fig:11} ARCA Bit Error Rate as function of power level of the receiver for the optical module data channel onshore. Black line is drawn just to guide the eye.}
   \end{minipage}\hfill
   \begin{minipage}{0.48\textwidth}
     \centering
     \includegraphics[width=1\linewidth]{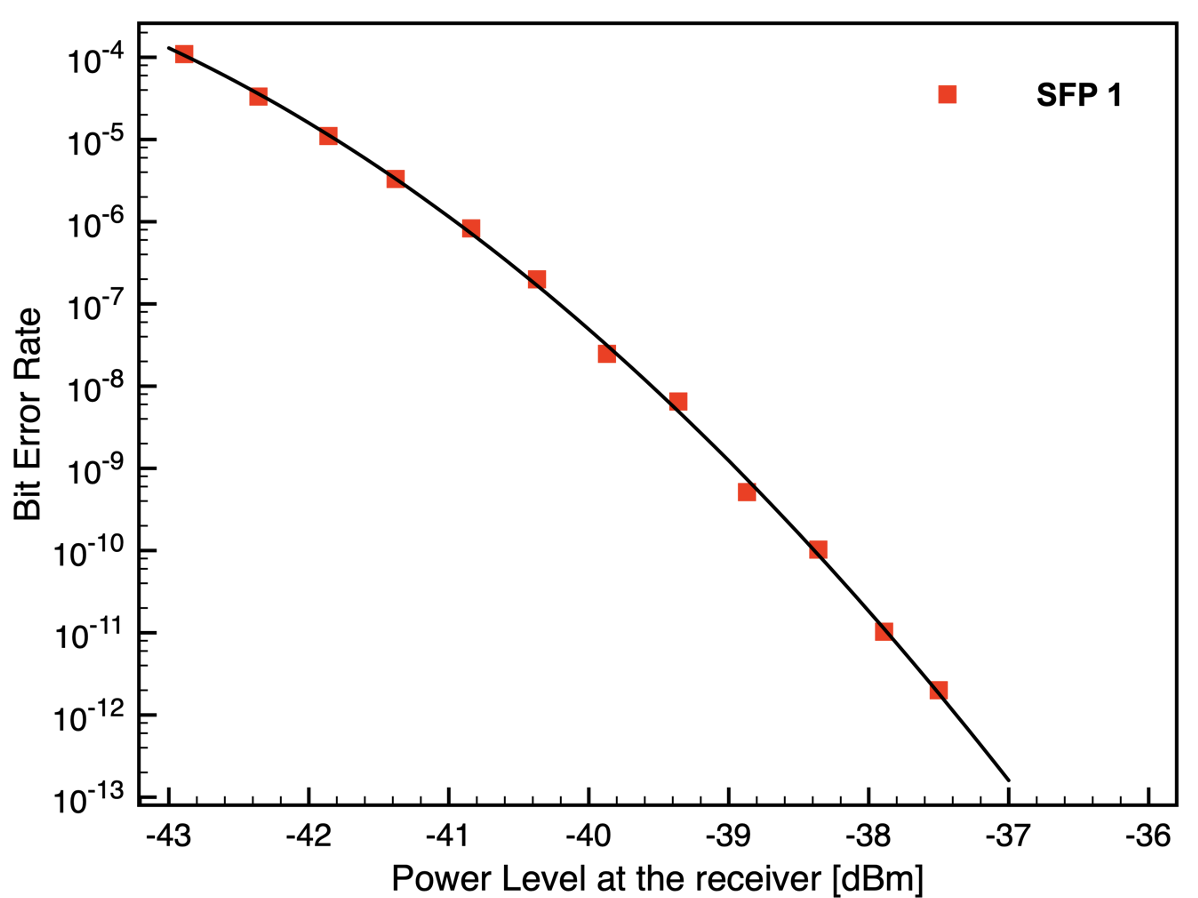}
     \caption{\label{fig:12} ORCA Bit Error Rate as function of power level of the receiver for the optical module data channel onshore. Black line is drawn just to guide the eye.}
   \end{minipage}
\end{figure}

The optical path of data from the optical modules was measured onshore at point \textbf{B} in Fig.~\ref{fig:4} and Fig.~\ref{fig:5}. Random DOM channels were chosen for this test run. The results for ARCA are presented in Fig.~\ref{fig:11} and show that there is only a weak dependence of the BER curves on the optical module channel. Therefore, for ORCA, only one channel is tested (Fig.~\ref{fig:12}).

\subsection{The optical budget}
The optical budget calculations performed during the system design phase were confirmed in the validation phase. The operational optical system margins for slow control, base module channel and optical modules data paths are calculated as the difference of the optical levels measured in the detector during the commissioning of the first deployed detection unit and the sensitivities shown in Fig.~\ref{fig:7} to Fig.~\ref{fig:12}. The levels measured during the commissioning include the gain provided by the amplifiers, and are the optical power level at the beginning of life before starting the ageing process and therefore the deterioration of the signal. For the ARCA system typical calculated margins are:

- about 20 dB for the slow control path;

- on average 20 dB for base module paths;

- on average 22 dB using RM-2 amplification and 28 dB using the ED-S amplification for the optical module data path.

\noindent The typical calculated margins for ORCA are:

- about 20 dB for the slow control path;

- on average 9 dB for the base module paths.

- on average 20 dB for the optical module data path.
The ORCA base module path margin is lower than the ARCA base module path due to the bypass of ED-F in the deployed system;
 
\noindent Since it is not possible to have a perfect replica of the deployed system, the margins calculated above might be overestimated by a few dBs. In addition, the measurements confirmed that, for some optical paths, the amplifier stages could be removed or temporarily bypassed. For the ARCA system, it was decided to remove the EDFA amplifier along the DOM data optical path in the junction box, and consequently extend the input dynamic range of the onshore preamplifier. As a result of the removal of the EDFA from the junction boxes the reliability of the offshore infrastructure of the ARCA detector improved. For the ORCA system it was decided to bypass ED-F in the shore station, since the beginning of life power levels before starting of the ageing process are well above the measured sensitivity.

\section{Next steps} 
As described in the introduction, when completed, the ARCA detector will comprise 230 detection units and ORCA will consist of 115 units. The implementation of the detectors has been subdivided into two smaller projects in order to adapt to the time lines of funding sources and procurement procedures and to allow for adaptation or replacement of obsolete technical parts. The first implementation phase foresees the construction of 32 detection units for ARCA and 48 for ORCA. The physical optical layer described in this paper is implemented in the detectors of the first phase with more than 500 optical modules in ARCA and more than 800 optical modules in ORCA. For the second phase a couple of adaptations are foreseen. The main reasons for the design change are scalability, reliability and standardisation. In order to finalise the second phase of KM3NeT, with three arrays of 115 detection units each, several technical requirements must be faced. The number of fibres required in the ARCA and ORCA networks has to be compliant with what are the market availability of standard underwater cables. The ratio between costs and number of fibres must be optimised. Moreover, the KM3NeT collaboration is currently validating the use of a fully standard White Rabbit technology for the communication between detector elements in the second phase of the project \cite{p}.
This solution requires the installation of White Rabbit switches in the base module of each detection unit. A high reliability switch board for offshore use was not available at the start of the design and implementation of the first phase of KM3NeT. The standard White Rabbit solution allows also for a minimisation of the number of fibers connecting the detection units to shore by using a single fibre for bidirectional communication. The solution simplifies both the network design and the time calibration procedures. With this new solution also the design of the hardware components in the shore station is simplified, resulting in a better scalability of rack units installed for each junction box in the control station.

\section{Concluding remarks}  
Details have been presented of the KM3NeT optical data transport system, which connects hundreds of optical modules and other components in the detector arrays in the deep sea with the control stations onshore. To comply with the scientific goals of KM3NeT, accurate time calibration between the many optical modules in the detector arrays is essential for the reconstruction of the neutrino events. For the first time in a deep sea neutrino telescope the White Rabbit protocol over Ethernet is used for the clock distribution. Downstream slow control and base module signals are broadcasted to all optical modules in the detector array. Synchronisation is achieved by communication between a White Rabbit master onshore and a White Rabbit slave unit inside the optical modules. Optical modules in the detector arrays are connected to the control station onshore via optical fibres in the detection unit and in the deep sea network. In order to verify the quality of the optical data transport system, BER values have been measured at various relevant points in a testbed that mimics the optical networks in the deep sea. Measured values indicate a high reliability of the networks. Moreover, data taking with the first detection units resulting in science papers demonstrates the good quality and reliability of the optical system \cite{q} \cite{r}.

\appendix

\acknowledgments
The authors acknowledge the Foton Institute for providing all the reports concerning the cited measurements performed at the their facilities in the years 2014-2015 during the Phase I architecture design.

The authors acknowledge the financial support of the funding agencies:
Agence Nationale de la Recherche (contract ANR-15-CE31-0020), Centre National de la Recherche Scientifique (CNRS), Commission Europ\'eenne (FEDER fund and Marie Curie Program), Institut Universitaire de France (IUF), LabEx UnivEarthS (ANR-10-LABX-0023 and ANR-18-IDEX-0001), Paris \^Ile-de-France Region, France;
Deutsche Forschungsgemeinschaft (DFG), Germany;
The General Secretariat of Research and Innovation (GSRI), Greece
Istituto Nazionale di Fisica Nucleare (INFN), Ministero dell'Universit\`a e della Ricerca (MIUR), PRIN 2017 program (Grant NAT-NET 2017W4HA7S) Italy;
Ministry of Higher Education, Scientific Research and Innovation, Morocco, and the Arab Fund for Economic and Social Development, Kuwait;
Nederlandse organisatie voor Wetenschappelijk Onderzoek (NWO), the Netherlands;
The National Science Centre, Poland (2021/41/N/ST2/01177);
National Authority for Scientific Research (ANCS), Romania;
Ministerio de Ciencia, Innovaci\'{o}n, Investigaci\'{o}n y Universidades (MCIU): Programa Estatal de Generaci\'{o}n de Conocimiento (refs. PGC2018-096663-B-C41, -A-C42, -B-C43, -B-C44 and refs. PID2021-124591NB-C41, -C42, -C43) (MCIU/FEDER, Generalitat Valenciana: Prometeo (PROMETEO/2020/019), Grisol\'{i}a (refs. GRISOLIA/2018/119, /2021/192) and GenT (refs. CIDEGENT/2018/034, /2019/043, /2020/049, /2021/023) programs, Junta de Andaluc\'{i}a (ref. A-FQM-053-UGR18), La Caixa Foundation (ref. LCF/BQ/IN17/11620019), EU: MSC program (ref. 101025085), Spain; María Zambrano program within the framework of grants for retaining in the Spanish university system (Spanish Ministry of Universities, funded by the European Union, NextGenerationEU).



\end{document}